\documentclass[12pt,a4paper]{article}
\usepackage[latin1]{inputenc}
\usepackage[english]{babel}
\usepackage{ifpdf}
\usepackage{epstopdf}
\usepackage{amsmath}
\usepackage{amsfonts}
\usepackage{amssymb}
\usepackage{abstract}
\usepackage{graphicx}
\usepackage{authblk}
\usepackage{cite}
\usepackage[left=2cm,right=2cm,top=2cm,bottom=2cm]{geometry}
\begin{document}
\title{Quantum Phase Transition in Dimerized Spin-1/2 Chains}
\author[a]{Aparajita Das}
\author[a]{Sreeparna Bhadra}
\affil[a]{Acharya Prafulla Chandra College, New Barrackpore, India}  \author[b]{Sonali Saha\thanks{{\bf email:} sonali@sncwgs.ac.in}}
\affil[b]{Sarojini Naidu College for Women, Kolkata, India}
\date{\today}
\maketitle
\begin{abstract}
Quantum phase transition in dimerized antiferromagnetic Heisenberg spin chain has been studied. A staircase structure  in the variation of concurrence within strongly coupled pairs with that of external magnetic field has been observed indicating multiple critical points. Emergence of entanglement due to external magnetic field or magnetic entanglement is observed for weakly coupled spin pairs in the same dimer chain. Though closed dimerized isotropic XXX Heisenberg chains with different dimer strengths were mainly explored, analogous studies on open  chains as well as closed anisotropic (XX interaction) chains with tilted external magnetic field have also been studied.
\end{abstract}
\section{Introduction}
Quantum phase transition (QPT) is a phenomenon where the non-analyticity in the energy level gives rise to a situation where energy fluctuation at the critical point vanishes~\cite{Sachdev1999}. Entanglement in a quantum system may be considered as a witness of this phenomenon (see e.g.~\cite{Arnesen2001, Zhou2003}). On the other hand, entanglement being an important resource~\cite{Ekert1991, Bennett1993, Bennett1996, Nielsen2010} in quantum information processing, this phenomenon is important from the perspective of its physical realization too.   

Spin-1/2 chains are some of the simplest qubit systems considered in the literature (see e.g.~\cite{Kane1998, Loss1998, Burkard1999, Imamoglu1999}) for the implementation quantum processor. Again the case of a two spin system, interacting through an isotropic XXX Heisenberg Hamiltonian in presence of an external magnetic field, the entanglement of formation has been calculated by Nielsen \cite{Nielsen1998}, which  was further extended for  a spin chain under periodic boundary conditions with a systematic investigations on the variation of  entanglement between various sites due to that of an external magnetic field and temperature  (see \cite{Arnesen2001}). The pairwise entanglement vanishes even at a vanishingly small temperature if the magnetic field exceeds a critical  value which characterises quantum phase transition. The effect of magnetic entanglement on isotropic XY model and thermal entanglement for anisotropic XY model \cite{Wang2001} and with non zero external magnetic field the effect of anisotropy on the same system had also been studied \cite{Kamata2002}. The dynamical phase transition in anisotropic XY model was also been studied~\cite{Sen(De)2005}. 

The effect of non uniform magnetic field on thermal entanglement for the two qubit  system with isotropic Heisenberg interaction~\cite{Sun2003} and for a two-qubit Ising model \cite{Terzis2004} were studied. Gong and Su \cite{Gong2009} have investigated pairwise thermal entanglement in $S=\frac{1}{2}XY$ chain (including its thermodynamic limit $N\rightarrow \infty$) and compared their results, with mean field frame work. They have found that in the case of first model there exist a common $T_c$, independent of $B$ for both ferromagnetic and anti-ferromagnetic interactions. It had also been found that in such a spin system $T_C$ is independent of $B$ for AF interactions. The studies were further extended for Heisenberg model with next to nearest neighbour interaction \cite{Plekhanov2008} and existence of two types of QPT was noticed. Other possibilities like existence of tilted external field \cite{Karthik2007}, mixed chain with spin-$ 1/2 $ and spin-$ 1 $ elements \cite{Hao2007}, were also explored. Far from equilibrium QPT has been observed in an open Heisenberg XY spin-$\frac{1}{2}$ chain characterized by a sudden appearance of long-range magnetic order in the non-equilibrium steady state as the magnetic field is reduced \cite{Prosen2007}. In fact  for last one and half decade or so, QPT in Ising chains and Heisenberg chains was extensively studied by different workers, only some of which we mentioned here.

Though the dimerized versions of Heisenberg  spin chains were studied since a long time back, they got renewed interests from the perspective of quantum communication.  On the other hand, for quite a long time spin chains are proposed candidates for communication medium between two quantum processors (see e.g.~\cite{Bose2007} and references therein).  But very recently a scheme of information transfer has been proposed~\cite{Yang2011} using dimerized spin-1/2 XXX model. Another scheme to generate end to end entanglement has also been proposed~\cite{Rafiee2013} using dimerized XXX and XXZ model. Although phase transition in dimerized spin-1/2 chain was studied in connection with transition betweeen gapped and ungapped energy spectra (see e.g.~\cite{Hida1992, Chitra1995}), quantum phase transition for the dimerized cases using thermal and magnetic entanglement has not been studied yet according to the authors' knowledge. The present paper explores this field for spin-1/2 dimerized Heisenberg chain.

\section{Formalism}
The exchange Hamiltonian of a  Heisenberg spin chain may be generalised as \begin{equation}
H_E=\sum_i (J_x \sigma_i^x\sigma_{i+1}^x+J_y \sigma_i^y\sigma_{i+1}^y+J_z \sigma_i^z\sigma_{i+1}^z),\label{exch}
\end{equation}
where $J_x$, $J_y$ and $J_z$ are the coupling strengths along three orthogonal directions. In addition if some external magnetic field $ \vec{B} $ is applied on the system, there will be additional on-site energy terms, generally called as Zeeman energy terms, as \begin{equation}
H_B=\sum_i\vec{B}.\vec{\sigma}_i.\label{zeeman}
\end{equation}
\begin{figure}[bh]
\centering
\includegraphics[scale=0.8]{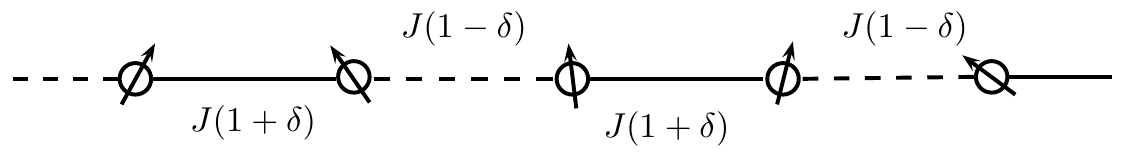}
\caption{\label{dimer} Schematic diagram of a dimer chain}
\end{figure}

The dimerized Hamiltonian for a chain of $ N $ spins in presence of an external uniform magnetic field acting along $ z $-axis is given by
\begin{equation}
H=B\sum_{i=1}^{N}\sigma_{i}^{z} +J\sum_{i=1}^{N}[1+(-1)^{i+1}\delta]\vec{\sigma}_i.{\vec\sigma}_{i+1}\label{dimH}
\end{equation} 	
where \[\vec{\sigma}_{i}=(\sigma^{x}_{i},\sigma^{y}_{i},\sigma^{z}_{i})\] are the Pauli spin matrices and the dimer strength is denoted by $ \delta $; with $ 0\le\delta\le 1 $. It is quite obvious from the expression that the coupling strength becomes strong and weak alternately for consecutive pairs which characterises a dimer. For the AFM chains the interaction strength $J>0$ and for the FM cases $J<0$. For a closed chain $ i $ runs from $ 1 $ to $ N $  in the second summation within the Hamiltonian expression, eq.~\eqref{dimH}, with the boundary condition $\vec{\sigma}_{N+1}\equiv\vec{\sigma}_1 $; while for open chain $ i $ simply runs from $ 1 $ to $ N-1 $. For an external magnetic field tilted by an angle $ \theta $ with $ z $-axis, eq.~\eqref{dimH} may be modified as \begin{equation}
H=B\cos\theta\sum_{i=1}^{N}\sigma_{i}^{z}+B\sin\theta\sum_{i=1}^{N}\sigma_{i}^{x}+J\sum_{i=1}^{N}[1+(-1)^{i+1}\delta]\vec{\sigma}_i.{\vec\sigma}_{i+1}.\label{tiltH}
\end{equation}

The system is taken to be in thermal state i.e., the state of the system when it is at thermal equilibrium with a thermal  bath (environment) at temperature $ T $, such that the density matrix, $\rho(T)=\exp{\left(-\frac{H}{kT}\right)}/{Z}$ where $Z$ is the partition function and $k$ is Boltzman's constant. 
In this communication we are reporting the variation of pairwise thermal entanglement for the strongly coupled and weakly coupled  spin pairs separately with that of magnetic field as well as that of dimer strength $\delta$ for a closed AFM chain(along with the open ended ones) of length $ N $. 

For this purpose first we need to find out the pairwise entanglement of qubit pairs. This may be obtained from the reduced density matrix for that pair of qubit, evaluated by tracing out the contribution of other qubits (or in other words taking partial trace on those degrees of freedom) from the total density matrix $\rho(T)$ of the  system. Thus the reduced density matrix of the subsystem (spin pair) of our concern may be expressed as \[\rho_R=\mathrm{Tr_B}\left[\rho(T)\right],\] where the complementary part of the system is named as subsystem B.

Now in general $ \rho_R $ will be of mixed state. For such a system the suitable measure of entanglement may be considered as entanglement of formation which may be expressed, for a two qubit system,  as a monotonic function of a simpler variable named as concurrence~\cite{Wootters1998}.  From reduced density matrix the following product matrix was calculated
\begin{equation}
R=\rho_R(\sigma_y\otimes\sigma_y)\rho_R^*(\sigma_y\otimes\sigma_y)
\end{equation}
If we represent $\lambda_i$'s with $i=1,2,3,4$ as the square root of the eigenvalues of $R$ in descending order then concurrence can be expressed as 
\begin{equation}
C=\mathrm{max}(0,\lambda_1-\lambda_2-\lambda_3-\lambda_4)
\end{equation} 
In this method we have described the standard bases,$|{00}\rangle,|{01}\rangle,|{10}\rangle,|{11}\rangle$ must be used.
In this study we have calculated the  concurrence indicating the thermal entanglement between different spin pairs. 

\section{Multiple Transitions and Staircase Structure }

\begin{figure}[ht]
\centering
\includegraphics[scale=0.6]{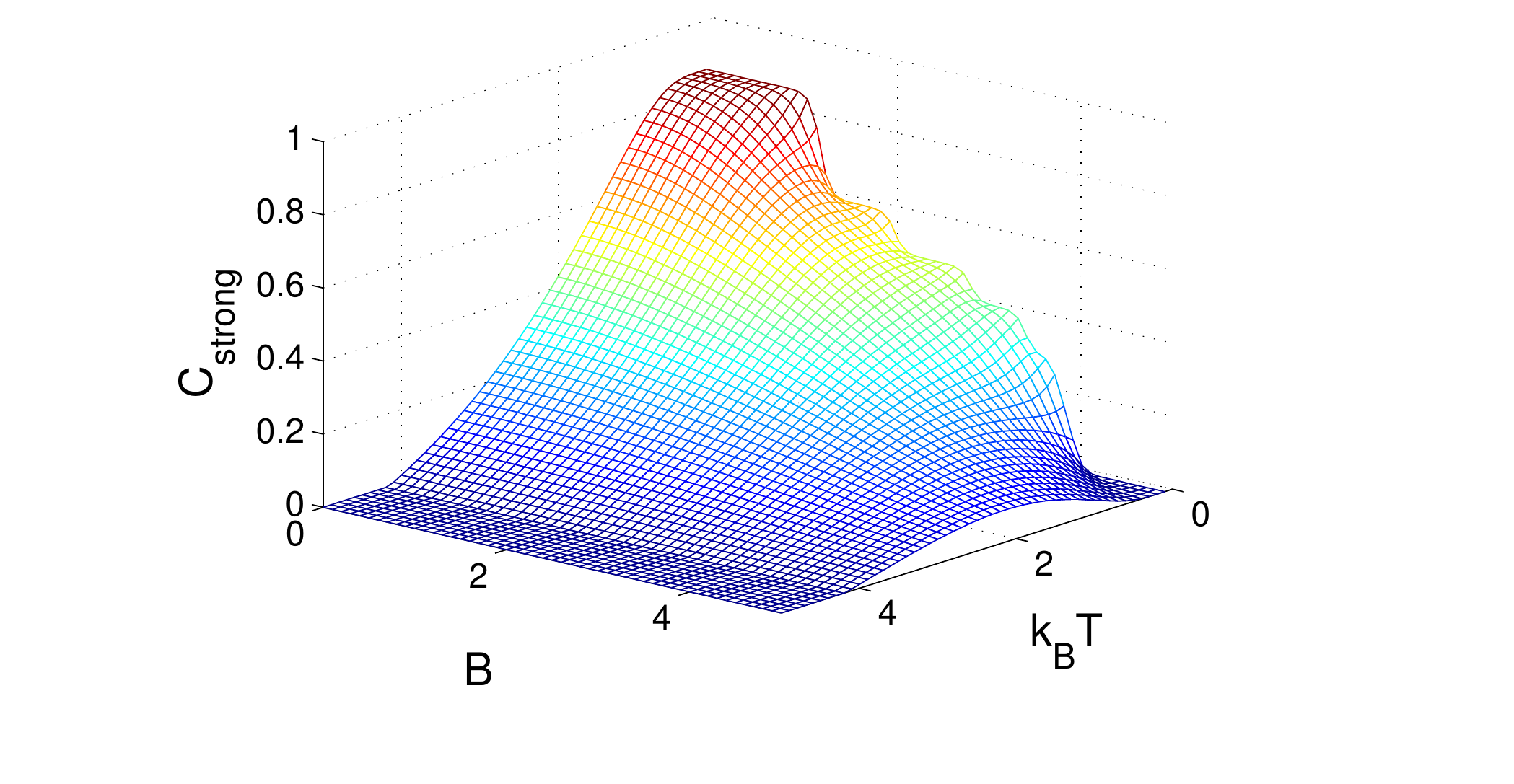}
\caption{\label{Fig.1} Variation of entanglement between strongly coupled spin pairs with temperature and external magnetic field for closed XXX chain. $ N=12, \delta= 0.2, \theta= 0$}
\end{figure}
Following the same line of investigation as done in the ref.~\cite{Arnesen2001}, we invetigated the pairwise entanglement of nearest neighbour pairs for a closed chain as quantified by the concurrence; but in this case for two types of pairs viz., strongly coupled and weakly coupled separately.

\begin{figure}[h!]
\centering
\includegraphics[scale=0.6]{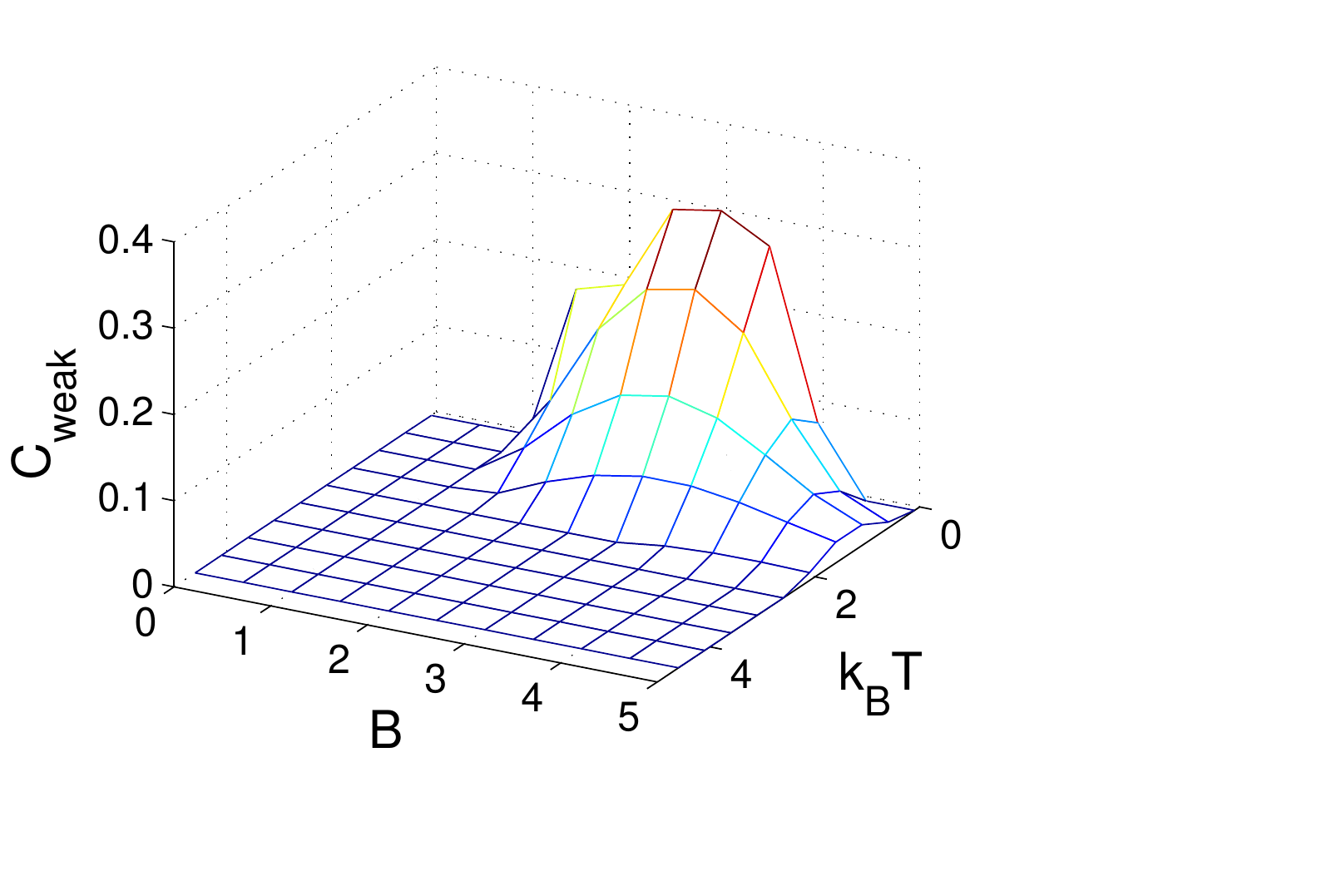}
\caption{\label{Fig.2} Variation of entanglement between weakly coupled spin pairs with temperature and external magnetic field for closed XXX chain.   $ N=12, \delta= 0.2, \theta= 0$.}
\end{figure}
It is quite apparent, from the figures\ref{Fig.1} and \ref{Fig.2} , that in the case of a dimer too, there is a sharp critical value of $ B =B_c$ across which the pairwise entanglement, both in the case of strongly coupled as well as of weakly coupled pairs abruptly vanishes. At very low temperatures the it is more abrupt due to the minimal presence of thermal fluctuations indicating some quantum phase transition at this value of $B$. The transition becomes progressively blurred at higher temperatures and ultimately vanishes at a critical temperature $T=T_c$ above which thermal fluctuations dominate destroying all pairwise entanglement in the chain. 
\begin{figure}[h!]
\centering
\includegraphics[scale=0.6]{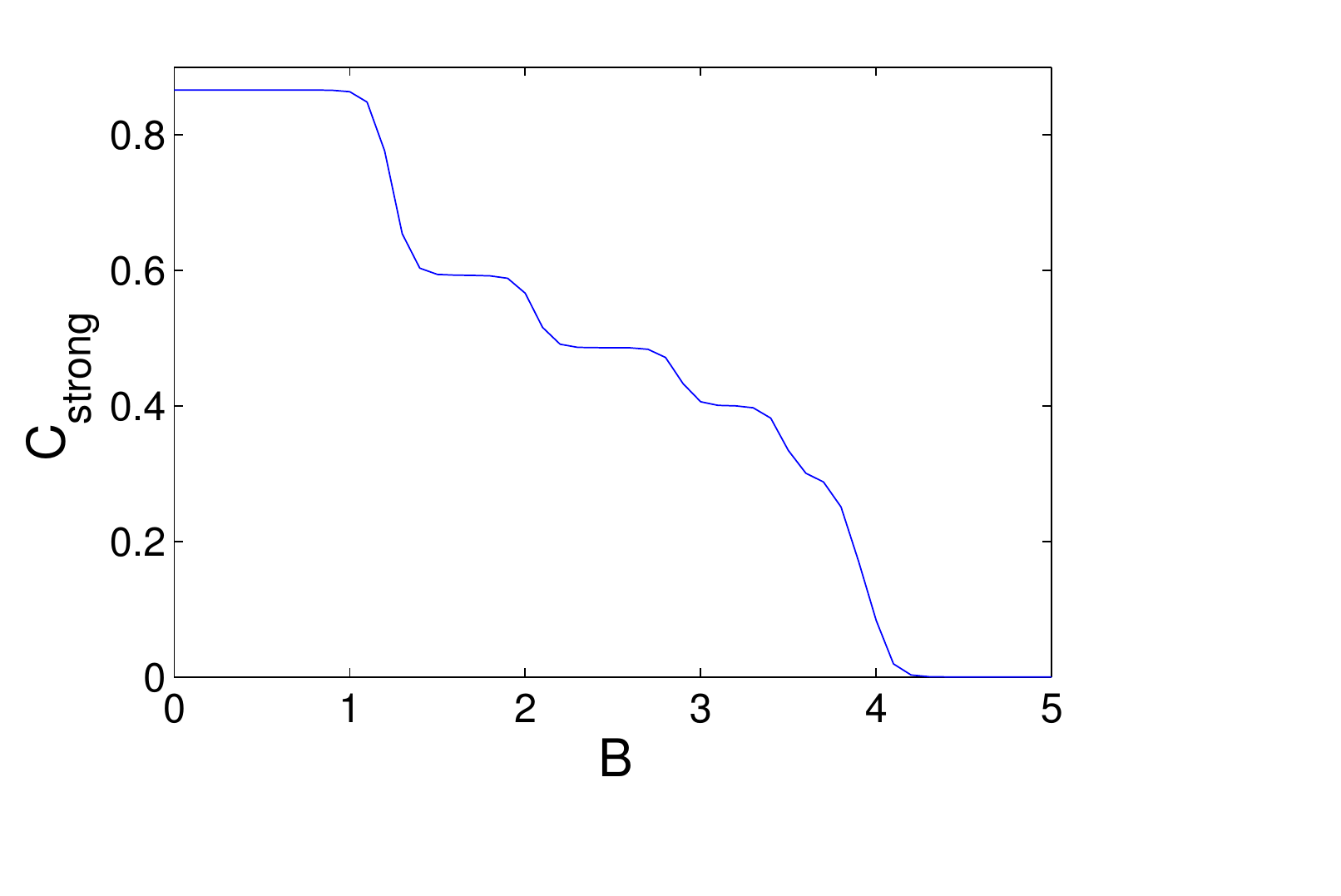}
\caption{\label{Fig.3(a)} Variation of entanglement within strongly coupled pair with magnetic field at constant temperature i.e with fixed $ k_BT=0.1 $. $ N=12, \delta= 0.2, \theta= 0$.}
\end{figure}
\begin{figure}[h!]
\centering
\includegraphics[scale=0.65]{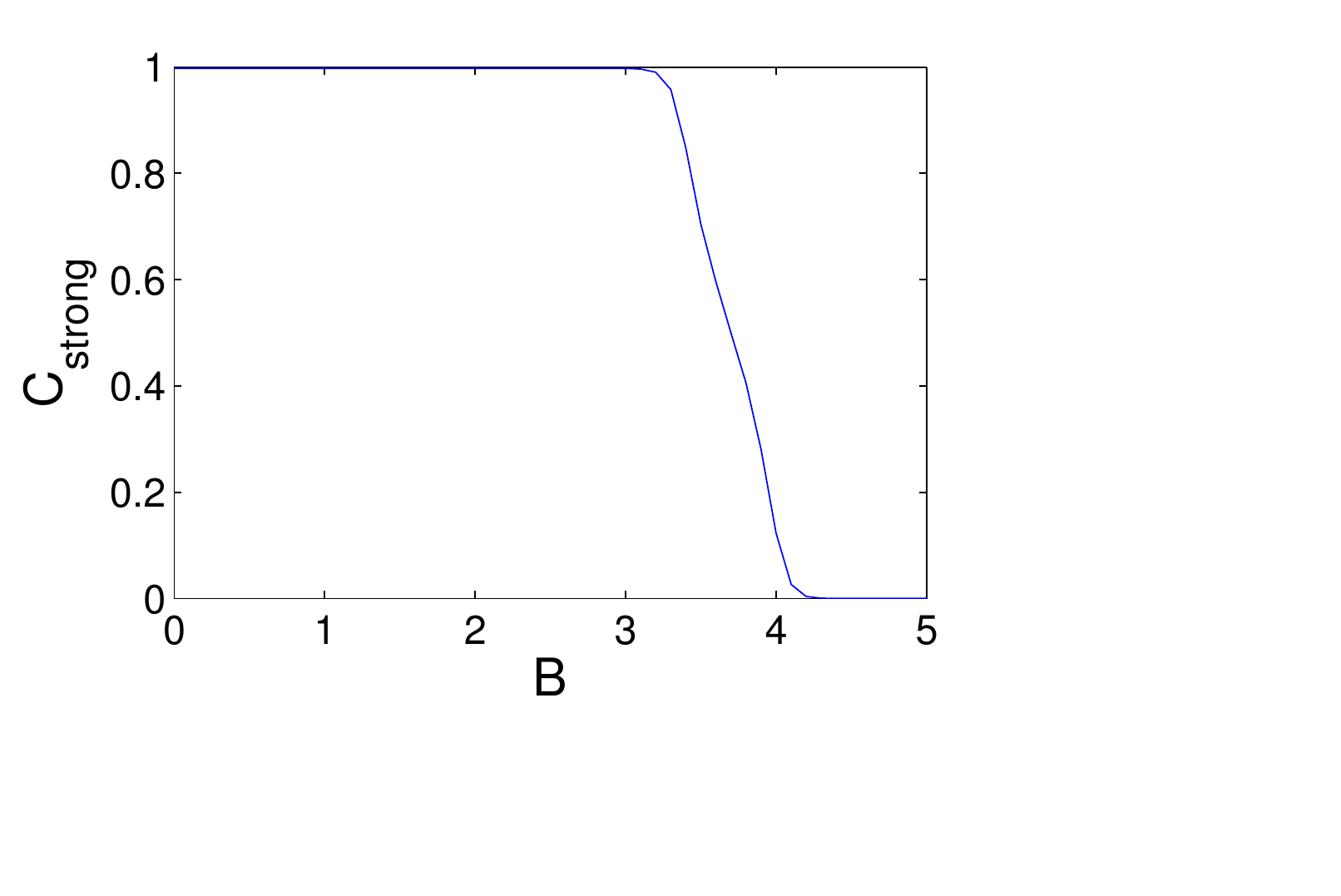}
\caption{\label{Fig.3(b)}  Variation of entanglement within strongly coupled pair with magnetic field at constant temperature i.e with fixed $ k_BT=0.1 $. $ N=12, \delta= 0.8, \theta= 0$.}
\end{figure}
But apart from these features also available in the non-dimerized version of Heisenberg chain as observed by several authors, there are certain distinctive characteristics available for the case of its dimerized version. First of all the values of concurrences in strongly coupled pairs are almost maximal ($ >0.9 $) while it is zero for weakly coupled pairs even for $ \delta =0.2 $ (see figs.~\ref{Fig.1} \& \ref{Fig.2}) in absence of any external magnetic field. Moreover particularly at low temperatures, there exist certain other discrete values of $B$, for each of which the pairwise concurrence for the strongly coupled pairs abruptly diminishes to a lower value and then remains steady with the increase of $B$, until next such critical value of $B$ arrives; giving rise to a staircase structure (see fig.~\ref{Fig.1}), which is easier to identify when one takes a constant temperature 2-d slice at a low temperature like the figure~\ref{Fig.3(a)} and \ref{Fig.3(b)}. It may be apparent from these figures (figs.~\ref{Fig.3(a)} and \ref{Fig.3(b)}) that this staircase structure is more pronounced with smaller values of dimer strength $ \delta $.
\begin{figure}[ht!]
\centering
\includegraphics[scale=0.6]{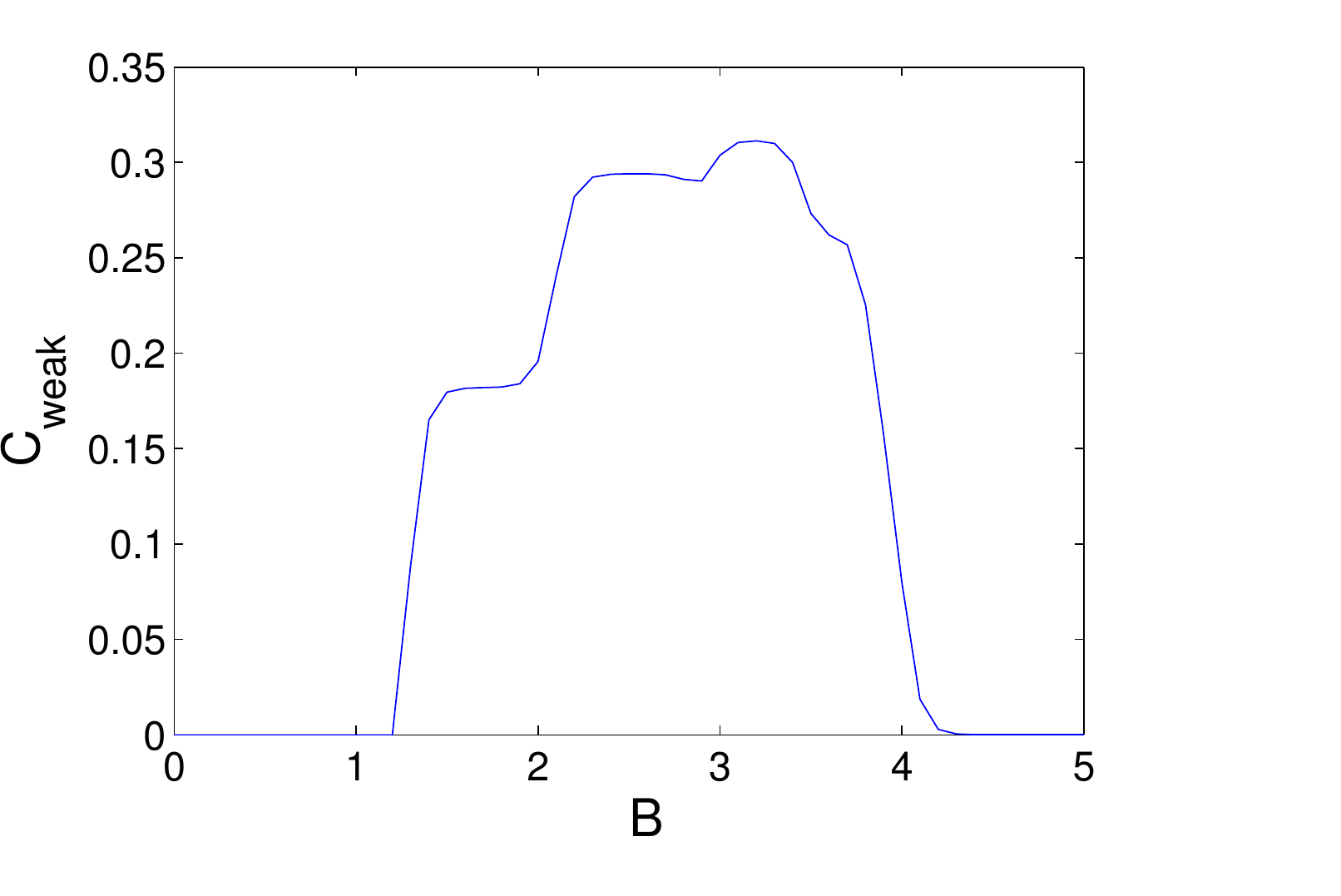}
\caption{\label{Fig.4(a)}Variation of entanglement within weakly coupled pair with magnetic field at constant temperature i.e with fixed $ k_BT=0.1 $. $ N=12, \delta= 0.2, \theta= 0$.}
\end{figure}

\begin{figure}[htb!]
\centering
\includegraphics[scale=0.8]{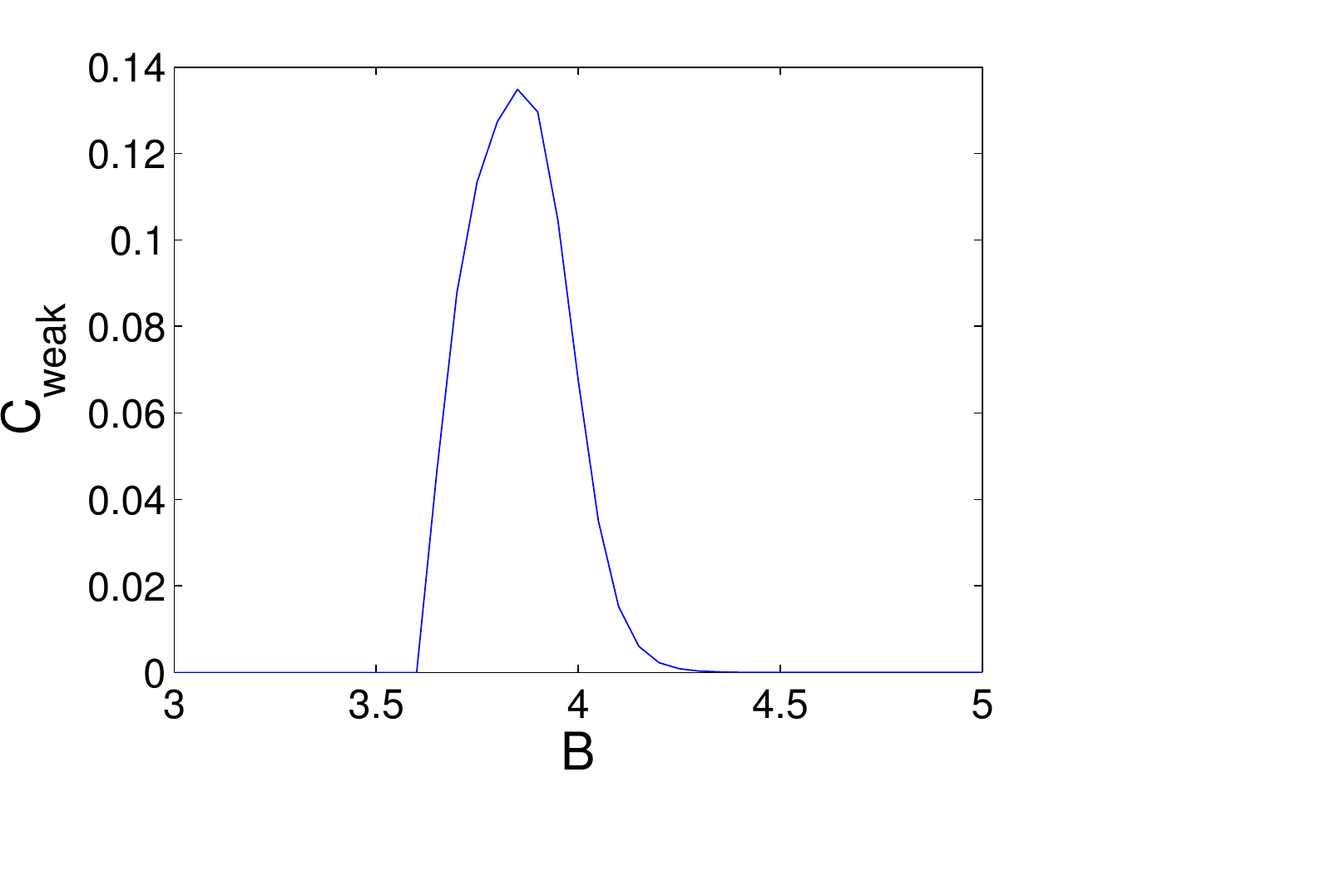}
\caption{\label{Fig.4(b)} Variation of entanglement within weakly coupled pair with magnetic field at constant temperature i.e with fixed $ k_BT=0.1 $. $ N=12, \delta= 0.8, \theta= 0$}
\end{figure}

The same observations for weakly coupled pairs of the same chain (fig.~\ref{Fig.2})  also revealed such staircase structure (particularly with small $ \delta $, as in fig.~\ref{Fig.4(a)}) but that structure becomes blurred with higher values of $ \delta $, except a certain surge near the final $B_c$ (see fig.~\ref{Fig.4(b)}). It may be noted that at the extreme value of $ \delta =1 $, the weakly coupled pairs get decoupled and hence entanglement within those pairs vanishes at that limit. Moreover it is apparent from the figs.~\ref{Fig.4(a)} \& \ref{Fig.4(b)} that at zero magnetic field weakly coupled pairs are completely separable and they get entangled  only after a specific value of external magnetic field sets in  (depending upon $ \delta $), which may be considered as a case of magnetic entanglement. 
\begin{figure}[htb]
\centering
\includegraphics[scale=0.55]{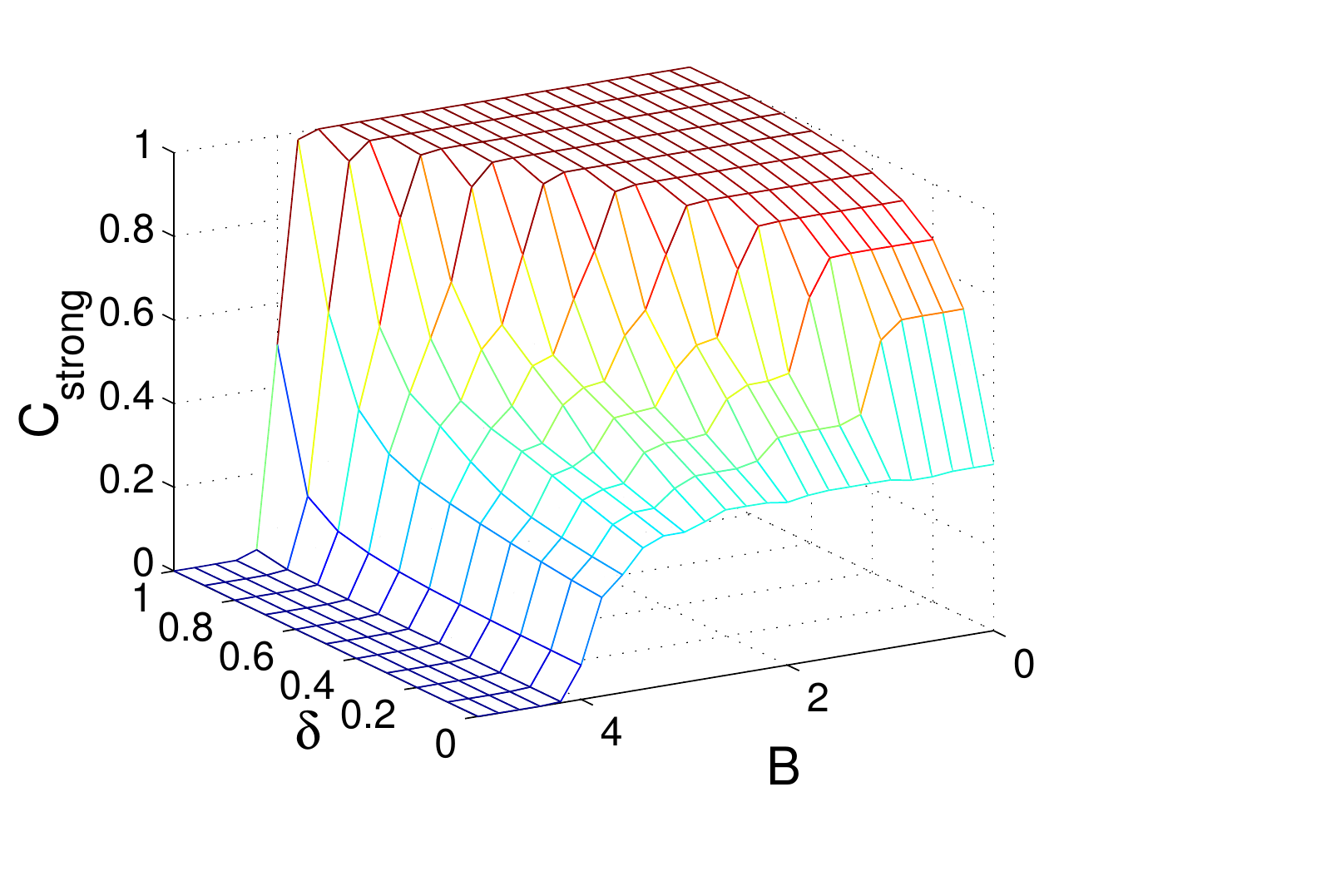}
\caption{\label{Fig.5} Variation of concurrence in strongly coupled pairs with the variation of $ B $ and $ \delta $ at a fixed temperature $ k_BT=0.1 $.}
\end{figure}

Next the variation  of concurrence within strongly coupled pairs with that of external magnetic field and that of dimer strength (see fig.\ref{Fig.5}) at a constant low value of temperature shows that this staircase feature is more predominant for relatively low values of dimer strengths.

From the 3-D plot (fig.`\ref{Fig.5}) it is clear that, if $ \delta $ increases from $ 0 $  even by a small amount (i.e. making it non-dimer chain) the staircase structure of the variation of concurrence with external magnetic field appears and the stair sizes vary with $ \delta $ and it vanishes when $ \delta $ approaches the value $ 1 $, the limit in which the chain breaks into  a collection of identical spin pairs interacting with XXX Hamiltonian.

The staircase  structure may be contributed by the gapped nature of the energy spectra for dimer chains, but further investigation is required to identify the actual mechanism.

\begin{figure}[htbp]
 \centering
 \includegraphics[scale=0.6]{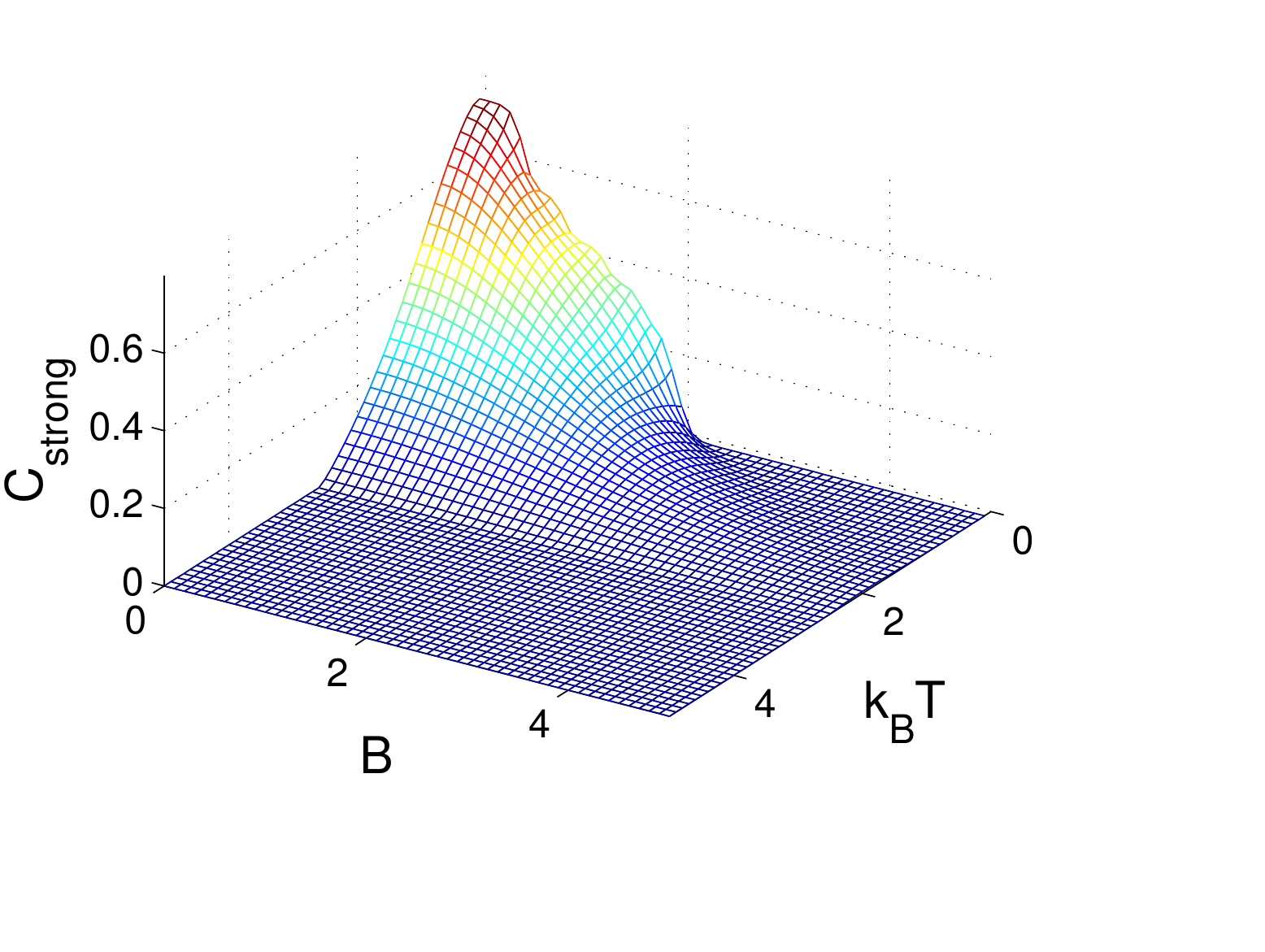}
 \caption{\label{Fig.6(a)} The variation of concurrence in strongly coupled pair [$ \delta =0.2 $] in XX model with $ B $ and $ k_BT $  for tilted $ B $ at an angle $ \theta = 0 $ with $ z$-axis.}
 \end{figure}

\begin{figure}[hbtp]
 \centering
 \includegraphics[scale=0.4]{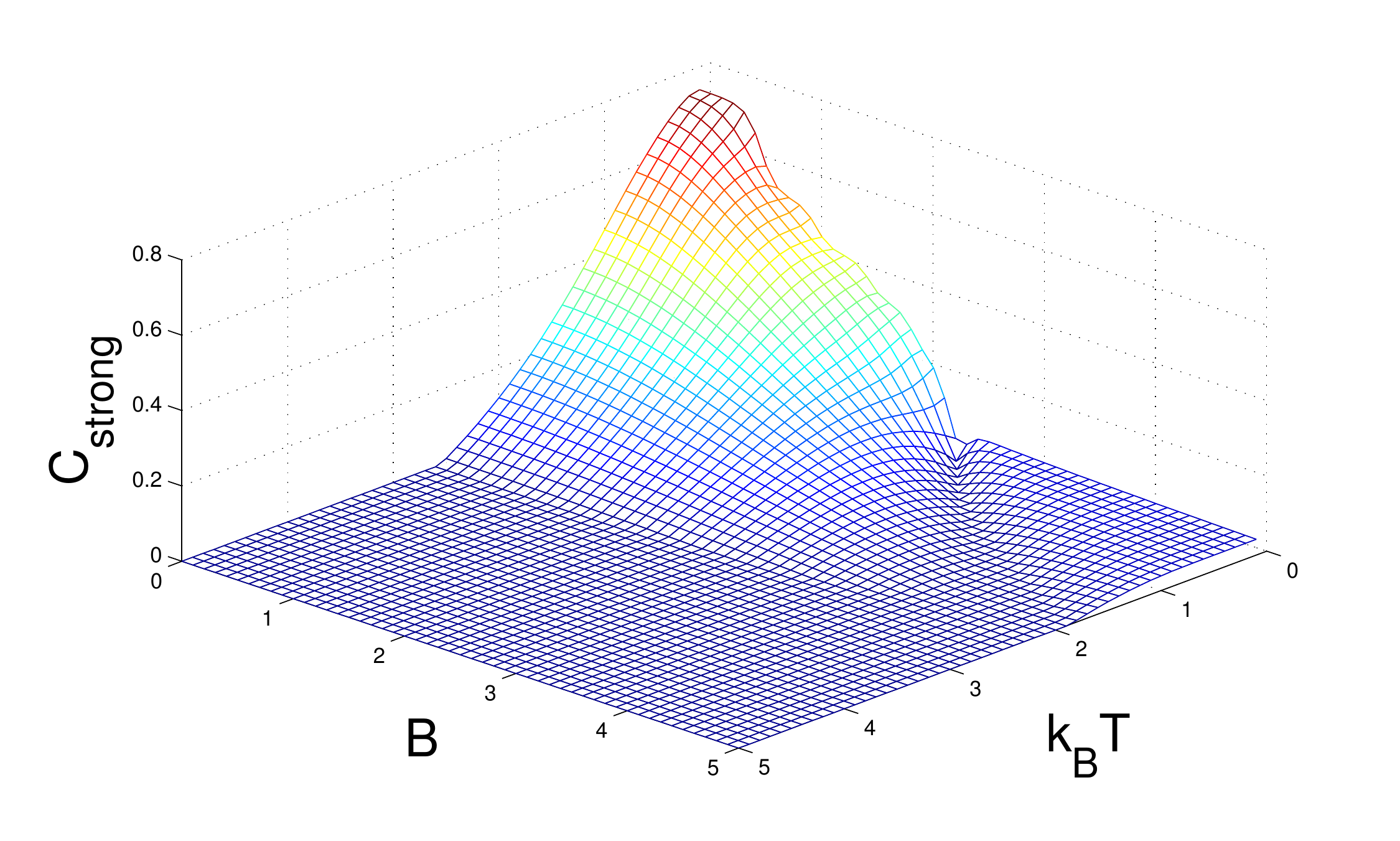}
 \caption{\label{Fig.6(b)} The variation of concurrence in strongly coupled pair [$ \delta =0.2 $] in XX model with $ B $ and $ k_BT $  for tilted $ B $ at an angle $ \theta =\pi/4 $ with $ z$-axis.}
 \end{figure}

\begin{figure}[hbtp]
 \centering
 \includegraphics[scale=0.55]{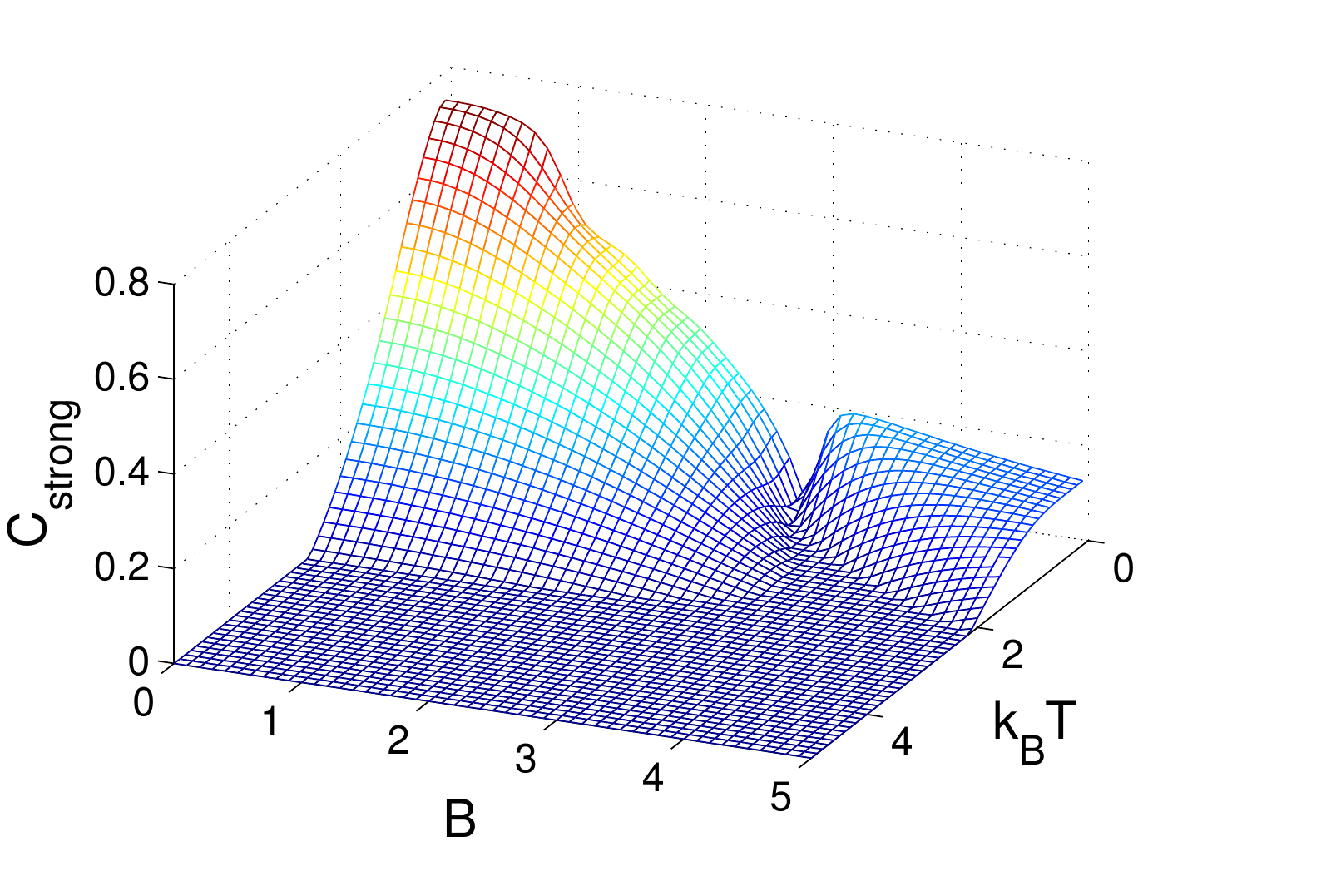}
 \caption{\label{Fig.6(c)} The variation of concurrence in strongly coupled pair [$ \delta =0.2 $] in XX model with $ B $ and $ k_BT $  for tilted $ B $ at an angle $ \theta =\pi/2 $ with $ z$-axis i.e. along $ x $-axis.}
 \end{figure}

To investigate the effect of inclination of external magnetic field, the dimerized XX chain is chosen.  The variation of the inclination of magnetic field has also been studied and for different angles the variations of the pairwise entanglement with field strength are plotted in figs.~\ref{Fig.6(a)}, \ref{Fig.6(b)} \& \ref{Fig.6(c)}. It has been found that for XX dimer chain even the strongly coupled pairs develop less amount of entanglement as expected intuitively, but it is observed that  at low temperature with the increase of $ B $, the entanglement first diminishes to almost zero at a critical value  of $ B $ and then again build up; moreover this build up increases with the relative strength of the longitudinal component of external field, i.e. $ B_x $ or in other word with the angle of tilt. Similar secondary build up of entanglement was also reported by some authors earlier for two qubit anisotropic Heisenberg chain with the variation of anisotropy (see~\cite{Kamata2002, Zhou2003}).


\begin{figure}[hbtp]
 \centering
 \includegraphics[scale=0.6]{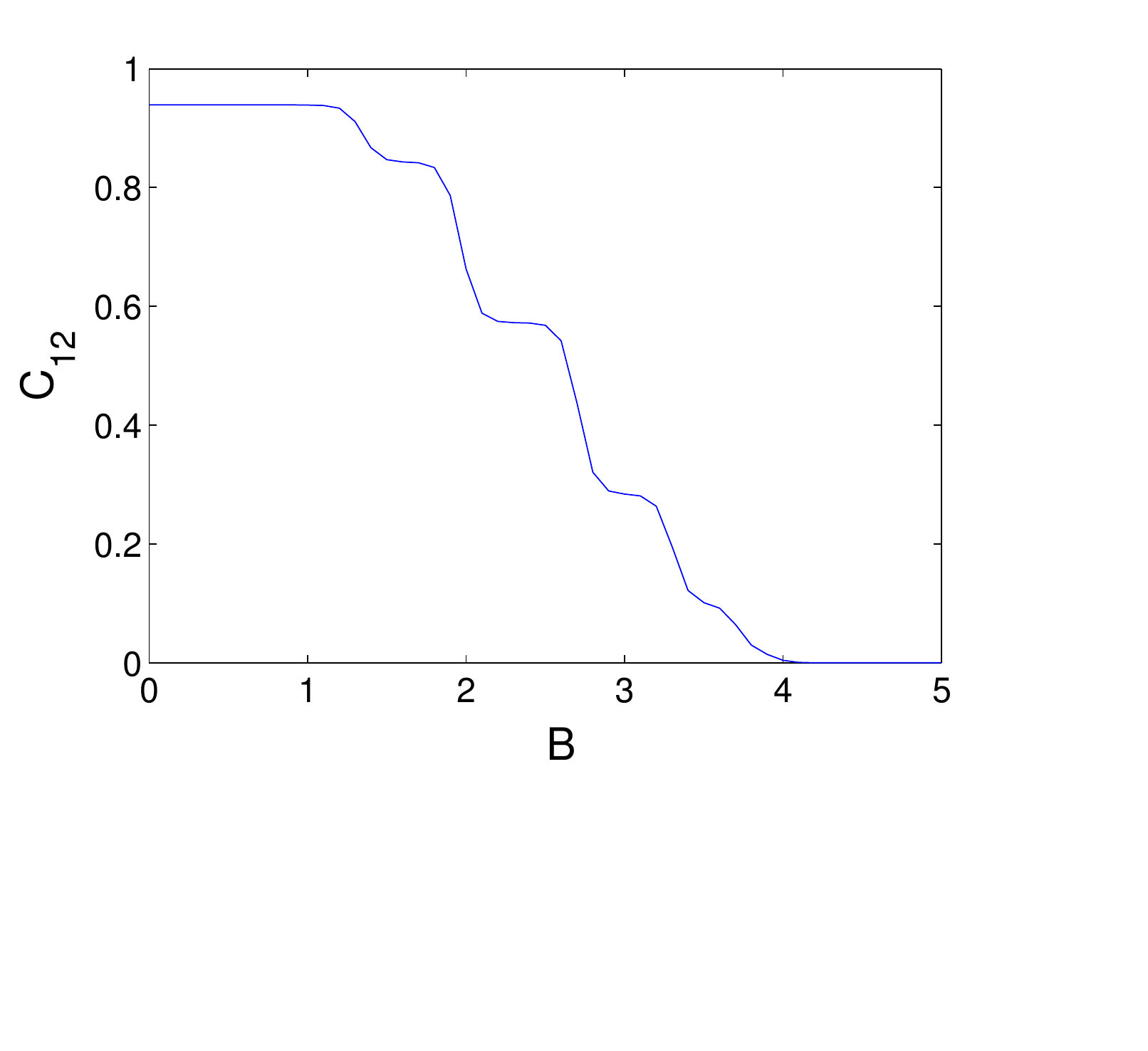}
 \caption{\label{Fig.7} Variation of concurrence in the first strongly coupled pair of an open XXX chain with external magnetic field $ B $ for $ \delta =0.2, N = 12 $}
 \end{figure}

\begin{figure}[h!]
 \centering
 \includegraphics[scale=0.6]{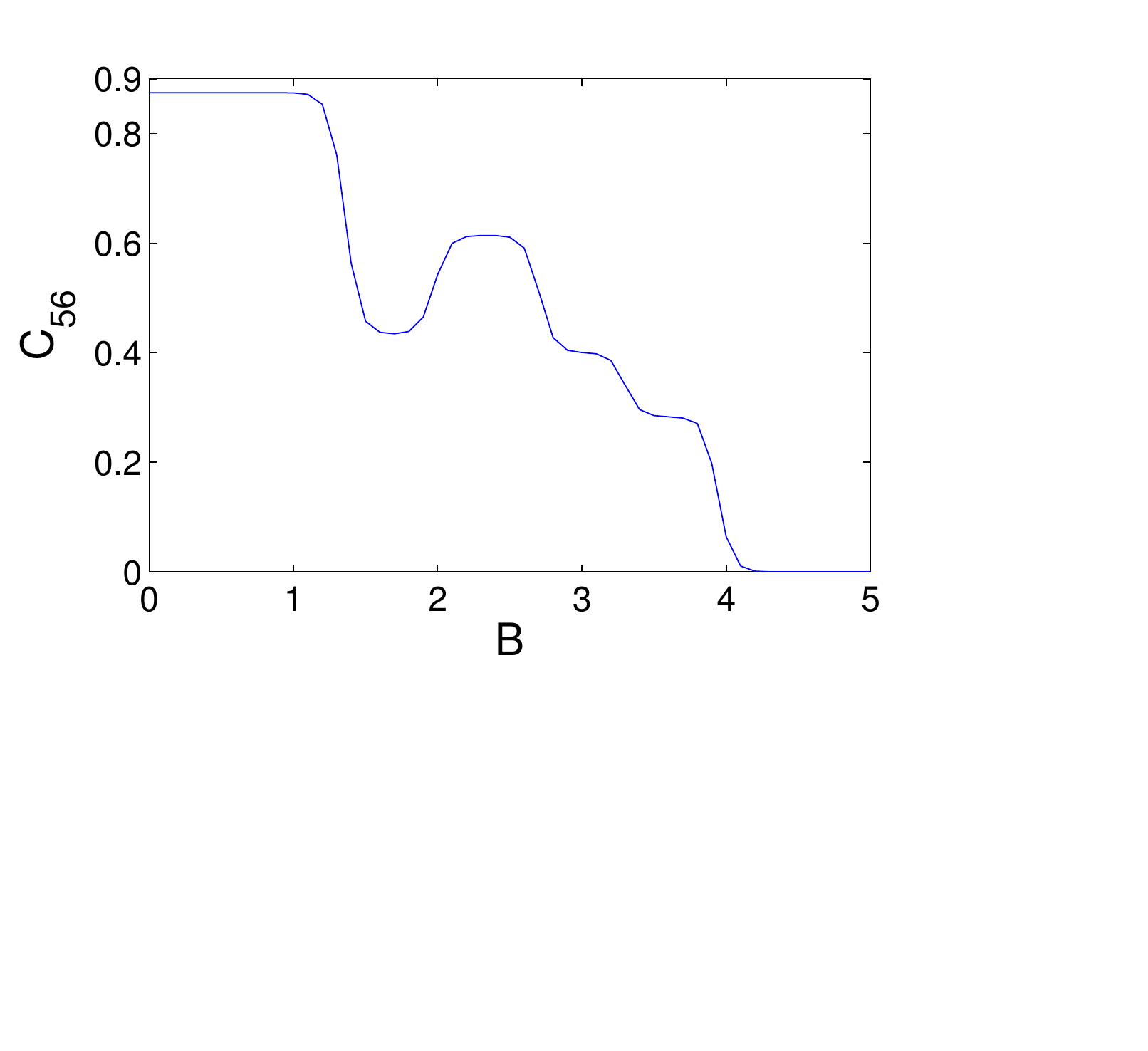}
 \caption{\label{Fig.8} Variation of concurrence in the fifth pair (strongly coupled) of an open XXX chain with external magnetic field $ B $ for $ \delta =0.2, N = 12 $}
 \end{figure}

 The studies already mentioned correspond to closed chains which have translational symmetry. But as this study has been motivated by the communication aspect of the dimers as recently proposed by some authors (refs), it is natural  to extend these studies for open XXX dimer chains. Due to absence of translational symmetry each pair has some unique feature in phase transition. We have chosen one strongly coupled one at the edge (fig.~\ref{Fig.7}) and another one near the middle of the chain (fig.~\ref{Fig.8}).
 
 It is clearly seen that the staircase structure is retained for the open chains too and in absence of external field strongly coupled ones are almost maximally entangled. The pair at the edge almost replicates (fig.~\ref{Fig.7}) the features corresponding to strongly coupled pairs in closed chains; but for the pair near the middle shows (fig.~\ref{Fig.8}) a build up at some intermediate values of $ B $, which is not found for strongly coupled pairs in closed XXX chains.
 
 \section*{Acknowledgement}
 AD and SB acknowledge the support offered by Acharya Prafulla Chandra College and Sarojini Naidu College for Women for carrying out this work part of which was done as part of their project work at Master's level. The work of SS was supported by the UGC MRP (Sanction No. F. PSW -- 164/13-14). SS further acknowledges the technical supports she got from Sankhasubhra Nag, SNCW, Kolkata. 

%\bibliography{quantum}

\begin{thebibliography}{10}

\bibitem{Sachdev1999}
S.~Sachdev.
\newblock {\em Quantum Phase Transitions}.
\newblock Cambridge UniversityPress, Cambridge, 1999.

\bibitem{Arnesen2001}
M.~C. Arnesen, S.~Bose, and V.~Vedral.
\newblock Natural thermal and magnetic entanglement in the 1d heisenberg model.
\newblock {\em Phys. Rev. Lett.}, 87(1):017901(4), 2001.

\bibitem{Zhou2003}
L.~Zhou, H.~S. Song, Y.~Q. Guo, and C.~Li.
\newblock Enhanced thermal entanglement in an anisotropic heisenberg xyz chain.
\newblock {\em Phy. Rev. A}, 68:024301(4), 2003.

\bibitem{Ekert1991}
Artur.~K. Ekert.
\newblock Quantum cryptography based on bell's theorem.
\newblock {\em Phys. Rev. Le}, 67:661, 1991.

\bibitem{Bennett1993}
Charles~H. Bennett, Gilles Brassard, Claude Crepeau, Richard Jozsa, Asher
  Peres, and William~K. Wootters.
\newblock Teleporting an unknown quantum state via dual classical and
  einstein-podolsky-rosen channels.
\newblock {\em Phys. Rev. L}, 70:1895, 1993.

\bibitem{Bennett1996}
C.~H. Bennett, D.~P. DiVincenzo, J.~A. Smolin, and W.~K. Wootters.
\newblock Mixed-state entanglement and quantum error correction.
\newblock {\em Phys. Rev. A}, 54:3824, 1996.

\bibitem{Nielsen2010}
Michael~A. Nielsen and Isaac~L. Chuang.
\newblock {\em Quantum Computation and Quantum Information}.
\newblock Cambridge University Press, Cambridge, 2010.

\bibitem{Kane1998}
B.~E. Kane.
\newblock A silicon-based nuclear spin quantum computer.
\newblock {\em Nature}, 393:133, 1998.

\bibitem{Loss1998}
Daniel Loss and David~P. DiVincenzo.
\newblock Quantum computation with quantum dots.
\newblock {\em Phys. Rev A}, 57:120--126, 1998.

\bibitem{Burkard1999}
Guido Burkard, Daniel Loss, and David~P. DiVincenzo.
\newblock Coupled quantum dots as quantum gates.
\newblock {\em Phys. Rev B}, 59:2070, 1999.

\bibitem{Imamoglu1999}
A.~Imamoglu, D.~D. Awschalom, G.~Burkard, D.~P. DiVincenzo, D.~Loss,
  M.~Sherwin, and A.~Small.
\newblock Quantum information processing using quantum dot spins and cavity
  qed.
\newblock {\em Phys. Rev. Lett}, 83:4204, 1999.

\bibitem{Nielsen1998}
M.A. Nielsen.
\newblock {\em Quantum Information Theory}.
\newblock PhD thesis, University of New Mexico, 1998.

\bibitem{Wang2001}
Xiaoguang Wang.
\newblock Entanglement in the quantum heisenberg xy model.
\newblock {\em Phys. Rev A}, 64:012313, 2001.

\bibitem{Kamata2002}
G.~Lagmago Kamta and Anthony~F. Starace.
\newblock Anisotropy and magnetic field effects on the entanglement of a two
  qubit heisenberg xy chain.
\newblock {\em Phys. Rev. Letts}, 88:107901, 2002.

\bibitem{Sen(De)2005}
Aditi Sen(De), Ujjwal Sen, and Maciej Lewenstein.
\newblock Dynamical phase transitions and temperature-induced quantum
  correlations in an infinite spin chain.
\newblock {\em Phys. Rev. A}, 72:052319, 2005.

\bibitem{Sun2003}
Y.~Sun, Y.~Chen, and H.~Chen.
\newblock Thermal entanglement in the two-qubit heisenberg xy model under a non
  uniform external magnetic field.
\newblock {\em Phys. Rev A}, 68:044301, 2003.

\bibitem{Terzis2004}
Andreas~F. Terzis and Emmanuel Paspalakis.
\newblock Entanglement in a two-qubit ising model under a site-dependent
  external magnetic field.
\newblock {\em Physics Letters A}, 333:438--445, 2004.

\bibitem{Gong2009}
Shou-Shu Gong and Gang Su.
\newblock Thermal entanglement in one-dimensional heisenberg quantum spin
  chains under magnetic fields.
\newblock {\em Phys. Rev A}, 80:012323--1--5, 2009.

\bibitem{Plekhanov2008}
E.~Plekhanov, A.~Avella, and F.~Mancini.
\newblock Entanglement in the 1d extended anisotropic heisenberg model.
\newblock {\em Physica B}, 403:1282?1283, 2008.

\bibitem{Karthik2007}
J.~Karthik, Auditya Sharma, and Arul Lakshminarayan.
\newblock Entanglement, avoided crossings, and quantum chaos in an ising model
  with a tilted magnetic field.
\newblock {\em Phys. Rev A}, 75:022304, 2007.

\bibitem{Hao2007}
Xiang Hao and Shiqun Zhu.
\newblock Entanglement in a quantum mixed-spin chain.
\newblock {\em Physics Letters A}, 366:206--210, 2007.

\bibitem{Prosen2007}
Tomaz Prosen and Iztok Pizorn.
\newblock Operator space entanglement entropy in a transverse ising chain.
\newblock {\em Phys. Rev. A}, 76:032316--1--5, 2007.

\bibitem{Bose2007}
Sougato Bose.
\newblock Quantum communication through spin chain dynamics: An introductory
  overview.
\newblock {\em Contemporary Physics}, 48:13, February 2007.

\bibitem{Yang2011}
S.~Yang, A.~Bayat, and S.~Bose.
\newblock Entanglement-enhanced information transfer through strongly
  correlated systems and its application to optical lattices.
\newblock {\em Phys. Rev A}, 84:020302(R), 2011.

\bibitem{Rafiee2013}
Morteza Rafiee and Hossein Mokhtari.
\newblock Long-distance entanglement generation by local rotational protocols
  in spin chains.
\newblock {\em Phys. Rev.}, 87:022304--1--6, 2013.

\bibitem{Hida1992}
Kazuo Hida.
\newblock Crossover between the haldane-gap phase and the dimer phase in the
  spin-1/2 alternating heisenberg chain.
\newblock {\em Phys. Rev. B}, 45:2207-- 2212, 1992.

\bibitem{Chitra1995}
R.~Chitra, Swapan Pati, H.~R. Krishnamurthy, Diptiman Sen, and S.~Ramasesha.
\newblock Density-matrix renormalization-group studies of the spin-1/2
  heisenberg system with dimerization and frustration.
\newblock {\em Phy. Rev. B}, 52:6581, 1995.

\bibitem{Wootters1998}
W.~K. Wootters.
\newblock Entanglement of formation of an arbitrary state of two qubits.
\newblock {\em Phys. Rev. Lett}, 80:2245, 1998.

\end{thebibliography}

\end{document}